\documentclass[twocolumn, amsmath, amssymb, aps]{revtex4-2}

\usepackage[dvipsnames]{xcolor}
\usepackage{graphicx}
\usepackage{enumitem}
\usepackage{siunitx}
\DeclareSIUnit\gauss{G}
\usepackage{booktabs}
\usepackage{multirow}
\usepackage{mathtools}

\usepackage[switch]{lineno}

\usepackage[utf8]{inputenc}

\newcommand\myshade{30}
\colorlet{mylinkcolor}{red}
\colorlet{mycitecolor}{orange}
\colorlet{myurlcolor}{orange}

\usepackage[%
  bookmarks=true,
  colorlinks,
  linkcolor=mylinkcolor!\myshade!black,
  urlcolor=myurlcolor!\myshade!black,
  citecolor=mycitecolor!\myshade!black,
  plainpages=false,
  pdfpagelabels,
  final,
  breaklinks=true
]{hyperref}

\hypersetup{
pdftitle={Entanglement-Enhanced Matter-Wave Interferometry in a High-Finesse Cavity},
pdfauthor={Thompson Lab},
pdfkeywords={interferometry, entanglement, quantum}
}

\newcommand{\omegaHF}{\omega_{\text{HF}}}
\newcommand{\omegaFSR}{\omega_{\text{FSR}}}

\newcommand{\OmegaTwoPh}{\Omega}
\newcommand{\Tevol}{\ensuremath{T_{\text{evol}}}}
\newcommand{\Tfall}{\ensuremath{T_{\text{fall}}}}

\newcommand{\thetaSQL}{\ensuremath{\theta_{\text{SQL}}}}
\newcommand{\thetaSQLi}{\ensuremath{\theta_{\text{SQL}}}}

\newcommand{\Nu}{N_{\uparrow}}

\newcommand{\Nd}{N_{\downarrow}}

\newcommand{\Nprime}{N}
\newcommand{\Ninit}{N_{0}}

\newcommand{\up}{\ensuremath{\ket{\uparrow}}}
\newcommand{\down}{\ensuremath{\ket{\downarrow}}}

\newcommand{\Jc}{J_{c}} 
\newcommand{\Js}{J_{s}} 

\newcommand{\deltaProbe}{\ensuremath{\delta_{c}}}
\newcommand{\deltaVS}{\ensuremath{\delta_{\text{vs}}}}
\newcommand{\deltaTwoPhoton}{\ensuremath{\delta}}
\newcommand{\deltaOAT}{\ensuremath{\delta_{p}}}

\newcommand{\Minc}{\ensuremath{M_{\text{i}}}}
\newcommand{\ntri}{\ensuremath{n_{\text{tri}}}}
\newcommand{\omegatri}{\ensuremath{\omega_{\text{tri}}}}

\newcommand{\alphanaught}{\ensuremath{\alpha_{0}}}
\newcommand{\TVS}{\ensuremath{t_{\text{vs}}}}
\newcommand{\Jzdiff}{\ensuremath{J_{zd}}}
\newcommand{\Coop}{\ensuremath{C}}
\newcommand{\hatVert}{\ensuremath{\hat{Z}}} 

\usepackage{physics}
\usepackage{braket}

\begin{document}
\title{Entanglement-Enhanced Matter-Wave Interferometry in a High-Finesse Cavity}

\author{Graham P. Greve}
 \thanks{These authors contributed equally.}
\author{Chengyi Luo}
 \thanks{These authors contributed equally.}

\author{Baochen Wu}

\author{James K. Thompson}
 \email{jkt@jila.colorado.edu}
 \affiliation{JILA, NIST and Department of Physics, University of Colorado, Boulder, CO, USA.}
\date{\today}

\begin{abstract}
Entanglement is a fundamental resource that allows quantum sensors to surpass the standard quantum limit set by the quantum collapse of independent atoms. Collective cavity-QED systems have succeeded in generating large amounts\cite{KasevichSqueezing2016, Thompson17dBSqueezing} of directly observed entanglement involving the \emph{internal} degrees of freedom of laser-cooled atomic ensembles \cite{VladanQNDClock2010, ThompsonVacuumRabiSplitting, PolzikStroboscopicSqueezing2015, Thompson10dBSqueezing, KasevichSqueezing2016, Thompson17dBSqueezing, ReichelSelfAmplifyingSqueezing2020, KasevichSqueezedClock2020, VladanEntangledClockLifetime2010, VladanOATExp2010, KasevichPhaseMagnification2016, VuleticSqueezedOpticalClock2020}. 
Here we demonstrate cavity-QED entanglement of \emph{external} degrees of freedom to realize a matter-wave interferometer of 700 atoms in which each individual atom falls freely under gravity and simultaneously traverses two paths through space while also entangled with the other atoms. We demonstrate both quantum non-demolition measurements and cavity-mediated spin interactions for generating squeezed momentum states with directly observed metrological gain $3.4^{+1.1}_{-0.9}$~dB and $2.5^{+0.6}_{-0.6}$~dB below the standard quantum limit respectively. An entangled state is for the first time successfully injected into a Mach-Zehnder light-pulse interferometer with $1.7^{+0.5}_{-0.5}$~dB of directly observed metrological enhancement. Reducing the fundamental quantum source of imprecision provides a new resource that can be exploited to directly enhance measurement precision, bandwidth, and accuracy or operate at reduced size. These results also open a new path for combining particle delocalization and entanglement for inertial sensors \cite{BordeSagnacInterferometer1991, Chu1999GravityAcceleration}, searches for new physics, particles, and fields \cite{hamilton2015DarkEnergy,TinoEquivalencePrincipleAI2017,MullerFineStructureAI,GuellatiKhelifaFineStructure2020,MAGIS100DarkMatterDescription2021}, future advanced gravitational wave detectors \cite{RajendranGravityWaveSensor2013,ThompsonGravWaveDetection2017}, and accessing beyond mean-field quantum many-body physics \cite{LevSpinOrbitCoupling2019, EsslingerSpinTexture2018, HemmerichTimeCrystal2021, Zimmermann2021Supersolid,PhysRevX.9.031009}.
\end{abstract}

\maketitle

Light-pulse matter-wave interferometers exploit the quantized momentum kick given to atoms during absorption and emission of light in order to split atomic wavepackets so that they traverse distinct spatial paths at the same time.  Additional momentum kicks then return the atoms to the same point in space to interfere the two matter-wave packets. The key to the precision of these devices is the encoding of information in the phase $\phi$ that appears in the superposition of the two quantum trajectories within the interferometer. This phase must be estimated from quantum measurements to extract the desired information. For $N$ atoms, the phase estimation is fundamentally limited by the independent quantum collapse of each atom to an rms angular uncertainty $\Delta \thetaSQL = 1/\sqrt{N}$~rad known as the standard quantum limit (SQL) \cite{WinelandProjectionNoise1993}. 

Here, we demonstrate for the first time a matter-wave interferometer \cite{ChuAI, PritchardAIReview2009} with a directly observed interferometric phase noise below the standard quantum limit, a result that combines two of the most striking features of quantum mechanics: the concept that a particle can appear to be in two places at once and entanglement between distinct particles. This work is also a harbinger of future quantum many-body simulations with cavities \cite{LevSpinOrbitCoupling2019, EsslingerSpinTexture2018, HemmerichTimeCrystal2021, Zimmermann2021Supersolid} that will explore beyond mean-field physics by directly modifying and probing quantum fluctuations, or in which the quantum measurement process induces a phase transition \cite{PhysRevX.9.031009}.

Quantum entanglement between the atoms allows the atoms to conspire together to reduce their total quantum noise relative to their total signal \cite{KitagawaUedaOAT, WinelandParam}.  Such entanglement has been generated between atoms using direct collisional \cite{OberthalerNumberSqueezing, OberthalerSqueezedBECInt, Schmiedmayer2011TwinAtom, ChapmanSqueezing2012, You2017TwinFock, KlemptBECSpatialEntanglement, TreutleinBECSpatialEntanglement} or Coulomb \cite{WinelandSixAtomCatState2005, BlattGHZState14Qubit2011} interactions, including relative atom number squeezing between matter-waves in spatially separated traps \cite{OberthalerNumberSqueezing, TreutleinBECSpatialEntanglement,Schmiedmayer2011TwinAtom} and mapping of internal entanglement onto the relative atom number in different momentum states\cite{KlemptMomentumEntanglementArxivXXX}.  A trapped matter-wave interferometer with relative number squeezing was realized in \cite{Schmiedmayer2011TwinAtom}, but the interferometer's phase was anti-squeezed and thus the phase resolution was above the standard quantum limit.

We demonstrate for the first time, the realization of cavity-QED entanglement generation between the external momentum states of different atoms using two distinct approaches that both rely on the strong collective coupling between the atoms and an optical cavity.  In the first approach, we realize cavity-enhanced quantum non-demolition (QND) measurements \cite{KasevichSqueezing2016, Thompson17dBSqueezing,ThompsonVacuumRabiSplitting,VladanQNDClock2010} to essentially measure and subtract out the quantum noise.  In the second approach, we utilize the cavity to mediate unitary interactions between the atoms to realize so-called one-axis twisting (OAT) \cite{KitagawaUedaOAT,VladanOATExp2010,VladanOATTheory2010,KasevichPhaseMagnification2016,VuleticSqueezedOpticalClock2020} or an all-to-all Ising interaction.  Both approaches have been realized for generating as much as 18.5~dB of entanglement \cite{Thompson17dBSqueezing, KasevichSqueezing2016}, but only between \emph{internal} states of atoms and with only the realization of directly observed enhancements in entangled microwave clocks \cite{KasevichSqueezedClock2020,VladanEntangledClockLifetime2010} and magnetometers \cite{Vuletic2019NearUnitary}. Cavity approaches to OAT \cite{HollandSqueezedBraggAI} and QND \cite{TinoMomSqueezingProposal} entanglement of purely Bragg interferometers have also been proposed.

\begin{figure*}[hbt!]
	\centering
	\includegraphics[width=\textwidth]{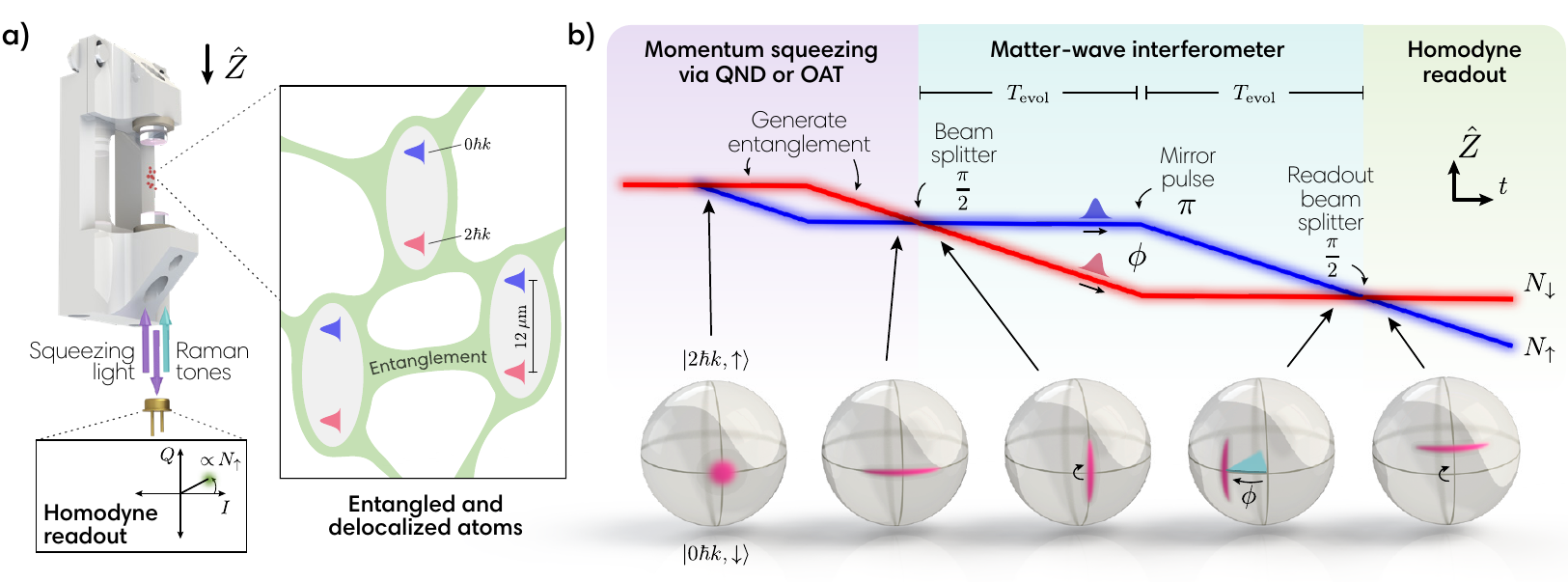}
	\caption{\textbf{Experimental overview.} \textbf{(a)} Ultracold atoms undergo guided free fall in a vertical high-finesse cavity. The atomic wavepackets are split and recombined by driving two-photon Raman transitions to provide quantized momentum kicks to the atoms. 
		\textbf{(right inset)} Intracavity atomic probe light generates entanglement between the atoms via either one-axis twisting dynamics or quantum non-demolition measurements made by \textbf{(bottom inset)} detecting the reflected atomic probe field's $Q$-quadrature with a homodyne detector \cite{ZilongsPRA, Thompson17dBSqueezing} . The entanglement between atoms is seen to persist over wavepacket separations exceeding $\SI{12}{\micro\meter}$.  \textbf{(b)} Space-time and Bloch sphere depictions of the generation and injection of the entanglement into a Mach-Zehnder matter-wave interferometer. Squeezing is first generated in the population basis, and then a Raman beam splitter pulse orients the squeezing for enhanced interferometer phase sensitivity. The two paths (red and blue) accrue a relative phase $\phi$ over time $2\,\Tevol$, the mirror pulse serves to re-overlap the wavepackets, and the readout beam splitter pulse creates interference that is read out as a population difference with sub-standard quantum limit sensitivity. Representative noise distributions are depicted on the Bloch sphere for various points in the interferometer.}
	\label{fig:setup}
\end{figure*}

Strong collective coupling to the cavity $N \Coop \gg1$ is the key requirement for both approaches to generate entanglement, where $\Coop$ is the single particle cooperativity parameter \cite{ZilongsPRA, VladanOATTheory2010, SorensenOATTheory2017}. Previously, an interferometer was operated in a low finesse cavity 
\cite{MullerCavityAI,MullerProbingGravity20Sec}, to provide power build-up, spatial mode filtering, and precise beam alignment. Here, we achieve matter-wave interferometric control \cite{ChuAI, PritchardAIReview2009} simultaneously with strong collective coupling $N \Coop \approx 500$ by operating inside a high cavity finesse $\mathcal{F} = 1.3\times10^5$ with small mode waist $w_0 = \SI{72}{\micro\meter}$.

Our two-mirror cavity is vertically-oriented along $\hatVert$ (Fig.~\ref{fig:setup}). The cavity has a power decay rate $\kappa = 2\pi \times \SI{56(3)}{~\kilo\Hz}$ at \SI{780}{\nano\meter}, mirror separation $L = \SI{2.2}{\centi\meter}$, and free spectral range  $\omegaFSR = 2 \pi \times \SI{6.7879}{\giga\Hz}$ (all error bars reported are 1$\sigma$ uncertainties). Rubidium atoms are laser cooled inside the cavity and then allowed to fall under gravity for a duration $\Tfall$, guided tightly along the cavity axis by a hollow (Laguerre-Gauss LG$_{01}$-like) blue-detuned optical dipole guide\cite{ThompsonHomogeneousCoupling} with thermal rms cloud transverse radius of $r_{\mathrm{rms}}= 4.7(8)~\mu$m $\ll w_0$ (see Methods).

\vspace{2mm}
\noindent\textbf{Manipulating matter-waves.} We manipulate matter-wave wavepackets using velocity-sensitive two-photon transitions with wavelength $\lambda=780$~nm.  The combined  absorption and stimulated emission of photons imparts $2 \hbar k$ momentum kicks oriented along the cavity axis, where $k=2 \pi/\lambda$ and $\hbar$  is the reduced Planck constant. 

\begin{figure*}[htb!]
	\centering
	\includegraphics[width=\textwidth]{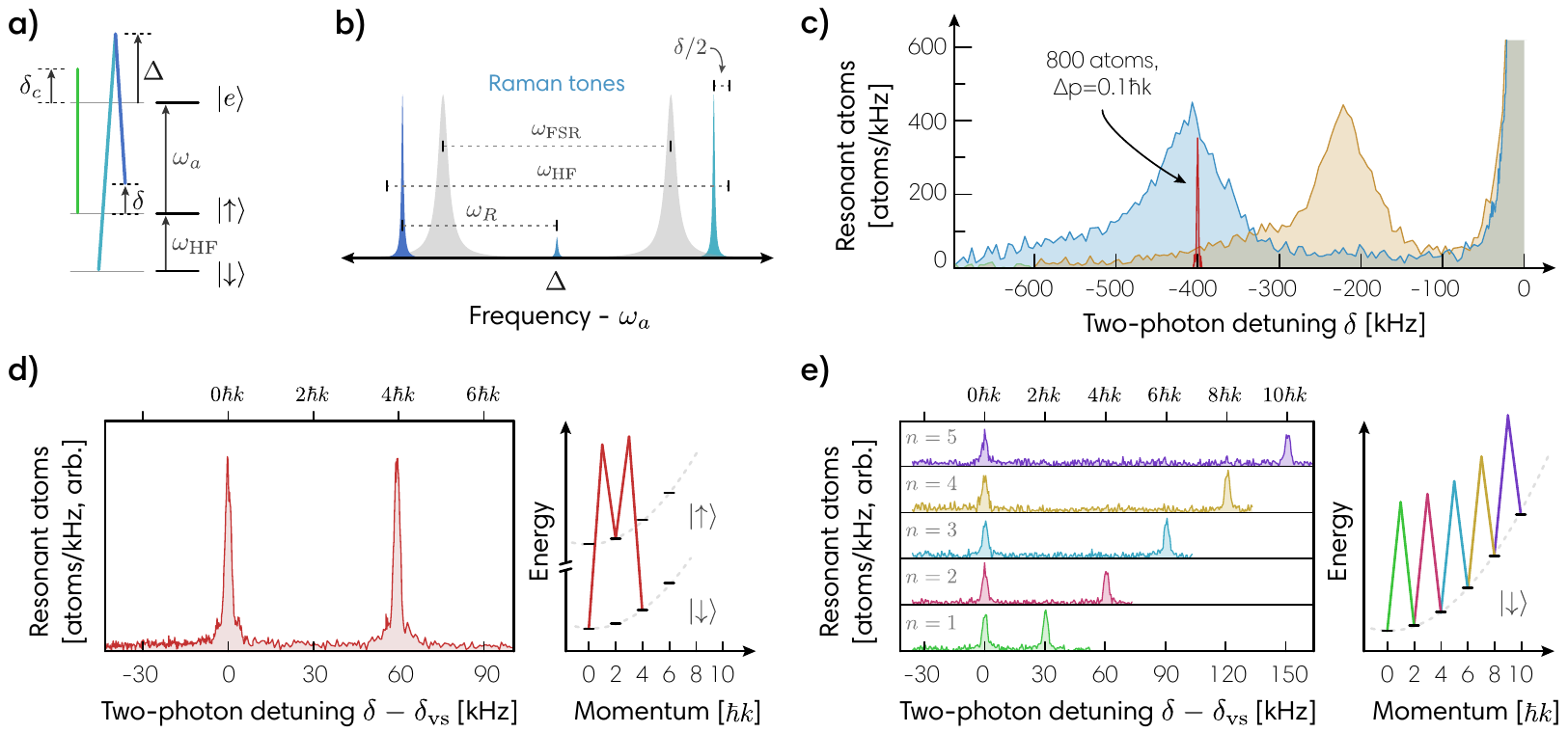}
	\caption{\textbf{Manipulating matter-waves in a high finesse cavity.}
		\textbf{(a)} Simplified energy-level diagram for $^{87}$Rb. The empty-cavity resonance used for probing (green) is detuned by $\deltaProbe$ from the $\up \rightarrow \ket{e}$ transition $\omega_a$. The Raman tones (blue) injected into the cavity drive a spin-changing $\up \leftrightarrow \down$ transition with two-photon detuning $\deltaTwoPhoton$ defined in a falling reference frame. \textbf{(b)} The Raman tones are derived from a laser detuned $\Delta$ from $\omega_a$, locked between two adjacent TEM$_{00}$ modes separated by $\omegaFSR$ (grey), and modulated at $\omega_R \sim \omegaHF/2$ for ground-state hyperfine splitting $\omegaHF$, leaving the tones detuned from the cavity resonances by $\pm 23$ MHz.
		\textbf{(c)} Atoms are prepared in $\down$ and allowed to fall for a duration $\Tfall = \SI{7.5}{\milli\second}$ (orange) or \SI{15}{\milli\second} (blue). The Raman coupling is applied at a fixed detuning $\delta$, after which the number of atoms in $\up$ is measured, revealing the axial velocity distribution. The full-width half-maximum of both distributions corresponds to a momentum spread of $5 \hbar k$, too broad for interferometry. During velocity selection, a group of about $800$ atoms with rms momentum spread $\Delta p = 0.1 \hbar k$ (red) are kept from the latter distribution while the rest are removed with transverse radiation pressure.  
		\textbf{(d)} After velocity selection at a two-photon detuning $\deltaVS$, a pair of Raman transitions can be used to place atoms into a superposition of $\ket{0\hbar k, \downarrow}$ and $\ket{ 4 \hbar k, \downarrow}$. Raman spectroscopy is used to verify the discrete velocity distribution.
		\textbf{(e)} Alternatively, Bragg transitions can be driven by adding amplitude modulation to the Raman tones. Here, a Bragg $\pi/2$ pulse splits the wavepacket, and consecutive $\pi$ pulses transfer additional momentum to create a superposition $\ket{0\hbar k, \downarrow}$ and $\ket{ 2 n \hbar k, \downarrow}$ with the momentum difference as large as $10 \hbar k$ shown here.
	}
	\label{fig:matter-waves}
\end{figure*}

For Raman transitions in which both momentum and spin states are changed, we utilize the magnetically-insensitive $^{87}$Rb clock states, $\down \equiv \ket{F=1,m_F=0}$ and $\up \equiv \ket{F=2,m_F=0}$, separated by the hyperfine transition frequency $\omegaHF \approx 2 \pi \times \SI{6.835}{\giga\Hz}$. The driving laser's frequency is stabilized between two TEM$_{00}$ longitudinal modes approximately $\Delta = 2 \pi \times \SI{85}{\giga\hertz}$ blue-detuned of $\up \rightarrow \ket{e}\equiv \ket{5^2\mathrm{P}_{3/2},F = 3}$ (Fig.~\ref{fig:matter-waves}(a)). As shown in Fig.~\ref{fig:matter-waves}(b), the cavity free spectral range is tuned such that two sidebands at $\pm \omega_R$ are approximately $\pm 2 \pi \times 23$~MHz from resonance with the closest TEM$_{00}$ mode when $2 \omega_R = \omega_{HF}$.  This  configuration allows enough light to nonresonantly enter the cavity for a two-photon Rabi frequency $\OmegaTwoPh = 2 \pi \times \SI{10}{\kilo\hertz}$.  By injecting the Raman tones non-resonantly and with opposite detunings, we greatly suppress laser frequency noise from being converted into phase and amplitude noise inside the cavity. Such noise manifests as noise in the Raman rotations and undesired Bragg scattering to other momentum states. The frequency difference of the sidebands is linearly ramped at rate \SI{25}{\kilo\hertz / \milli\second} to compensate for the acceleration of the atoms by gravity (see Methods).

In Fig.~\ref{fig:matter-waves}(c), we show the initial axial velocity spectrum of the atoms as mapped out by inducing velocity-dependent spin flips. We use this same process to select atoms within a narrow range of initial velocities for coherent manipulation of matter-waves, resulting in approximately $\Ninit = 800-1200$ atoms in $\down$ with rms momentum spread $\Delta p = 0.1 \hbar k$ set by choice of the two-photon Rabi frequency $\OmegaTwoPh = 2\pi \times \SI{1.4}{\kilo\hertz}$ (see Methods).

In Fig.~\ref{fig:matter-waves}(d) we demonstrate the quantized nature of the momentum kicks imparted by the intracavity Raman transitions.  After velocity selection, a $\pi/2$ pulse is followed by a second Raman $\pi$ pulse to place the atoms into a superposition of $\ket{ 0 \hbar k,\downarrow}$ and $\ket{4 \hbar k, \downarrow}$ in the falling frame of reference.  We observe this as two distinct peaks separated in the subsequent velocity spectrum. Future interferometers might evolve in such superpositions so as to minimize systematic errors and dephasing due to differential environmental couplings between $\up$ and $\down$.

Complementary to hyperfine spin-state changing Raman transitions, we also demonstrate intracavity Bragg transitions in this high finesse and high cooperativity cavity.  The Bragg coupling (see Methods) connects states $\ket{n \hbar k}\leftrightarrow\ket{(n+2) \hbar k}$ with no change in the spin degree of freedom, as shown in Fig.~\ref{fig:matter-waves}(e).  After velocity selection, the wavepacket is coherently split by a Bragg $\pi/2$ pulse, followed by successive $\pi$ pulses to transfer momentum to one of the wavepacket components for a momentum difference of up to $10\,\hbar k$.  Access to Bragg transitions opens the door to both large momentum transfer operations for greater sensitivity and to improved coherence times in future work.

\vspace{2mm}
\noindent\textbf{Squeezing on momentum states.}
We now turn our attention to creating entanglement between atoms that includes this \emph{external} degree of freedom. We describe the collective state of our matter-wave interferometer using a Bloch sphere with average Bloch vector $\vec{J} = \langle \hat{J}_x \hat{x} +\hat{J}_y \hat{y} +\hat{J}_z \hat{z} \rangle $ of length $J \equiv \left| \vec{J} \right| \le \Ninit/2$ in a fictitious coordinate space (Fig.~\ref{fig:setup}(b)). The collective pseudospin projection operators are defined as $\hat{J}_z \equiv \frac{1}{2} \left( \hat{\Nu}-\hat{\Nd} \right)$ with collective population projection operators $\hat{\Nu} = \sum_{i}^{\Ninit} \ket{a}_i \prescript{}{i}{\bra{a}}$ and $\hat{\Nd} = \sum_{i}^{\Ninit} \ket{b}_i \prescript{}{i}{\bra{b}}$, and similarly for other pseudospin projections, where $\ket{a}_i=\ket{2 \hbar k, \uparrow}_i$ and $\ket{b}_i = \ket{0 \hbar k, \downarrow}_i$ for the $i$th atom. We use a Raman $\pi/2$ pulse to nominally prepare all atoms in an unentangled coherent pseudospin state described by the Bloch vector $\vec{J}=J \hat{x}$.  The standard quantum limit arises from the non-zero variance of the spin projection operators $\left(\Delta J_z\right)^2 =  \langle \hat{J}_z^2\rangle - \langle \hat{J_z} \rangle^2 \ne 0$, etc. and is visualized on the Bloch sphere as a quasi-probability distribution of the orientation of the Bloch vector from trial to trial. We prepare squeezed momentum states using both quantum non-demolition measurements \cite{ZilongsPRA, KasevichSqueezing2016, Thompson17dBSqueezing} and one-axis twisting \cite{KitagawaUedaOAT, VladanOATExp2010, VladanOATTheory2010} in which the quantum noise is reduced in one spin-momentum projection at the expense of increased quantum noise along the orthogonal projection.

The Wineland parameter $W$ characterizes the phase enhancement of a squeezed state with phase uncertainty $\Delta\theta$ that is certified to arise from entanglement between the atoms \cite{WinelandParam},
	\begin{align}
		W = \left(\frac{\Delta\theta}{\Delta\thetaSQLi} \right)^2. \label{eq:Wres}
	\end{align}
Physically, it is the reduction in the angular noise variance of the phase estimation relative to the standard quantum limit $\Delta\thetaSQLi=1/\sqrt{\Nprime}$ one would have for a pure state with a Bloch vector length $\Jc=\Nprime/2$ equal to that of the actual mixed or partially decohered state prepared without the squeezing operation (see Methods) .

\begin{figure}[hbt!]
	\centering
	\includegraphics[width=3.375in]{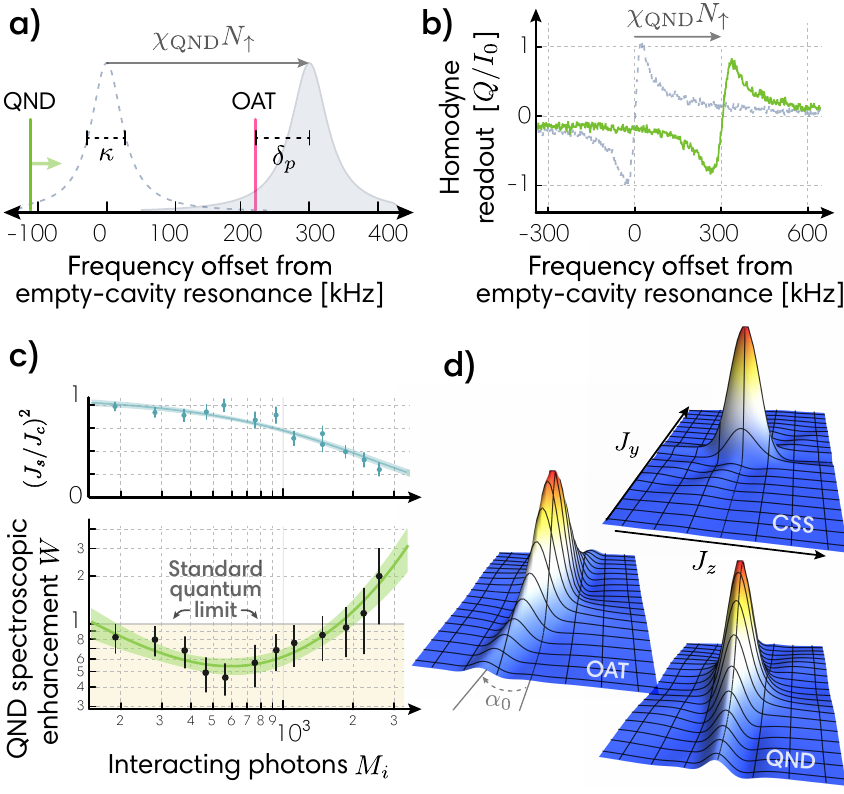}
	\caption{\textbf{Momentum squeezing via one-axis twisting and quantum non-demolition measurements.} \textbf{(a)} Probe frequency setup for OAT and QND measurements. During OAT, the laser is fixed at a detuning from cavity resonance $\deltaOAT$. QND measurements are made by sweeping the probe laser frequency over cavity resonance and detecting the $Q$ quadrature of the reflected field \cite{ZilongsPRA, Thompson17dBSqueezing}.  \textbf{(b)} QND probe sweeps measured in homodyne and normalized to the full reflected field on resonance $I_0$, shown for the empty cavity (gray) and for 900 atoms in $\up$ (green). The observed frequency shift allows us to measure the collective population operator $\hat{N}_{\uparrow}$ with measurement outcome $\Nu$, without knowing which atoms are in $\up$.  The probe is sweeping \SI{1.5}{\mega\hertz/\milli\second} and the atom-cavity detuning is $\deltaProbe = 2 \pi \times \SI{175}{\mega\hertz}$. Free space scattering of probe light results in a slight broadening and reduced amplitude of the observed signal \cite{ZilongsPRA}.  \textbf{(c)} QND measurements are used to pre-measure the quantum noise in the spin projection $J_z$ and subtract it from a final measurement as in \cite{Thompson17dBSqueezing}. Increasing the number of probe photons $M_i$ results in a more precise pre-measurement, but at too high of a photon number, free space scattering causes shortening of the Bloch vector (top) and spontaneous Raman scattering to other states. Squeezing is characterized by the spectroscopic enhancement $W$ (bottom) which reaches an optimum below the standard quantum limit at $M_i=600$ photons. Data is fit with 68\% confidence bands and all error bars reported are 1$\sigma$ uncertainties. \textbf{(d)} State tomography \cite{Thompson17dBSqueezing} was performed by applying a variable-duration pulse with rotation axis aligned with the Bloch vector to reconstruct the spin-momentum quasi-probability distributions in the $J_y-J_z$ plane for a coherent spin-state (CSS), a QND-squeezed state, and an OAT-squeezed state.}  
	\label{fig:squeezing}
\end{figure}

Collective QND measurements of the free falling atomic samples are used to estimate the number of  atoms in different spin-momentum states without revealing single-particle information \cite{ZilongsPRA, ThompsonHomogeneousCoupling}. The two momentum states interact differently with the optical cavity because they carry distinct spin labels. We tune a TEM$_{00}$ cavity mode with resonance frequency $\omega_c$ to the blue of the $\up \rightarrow \ket{e}$ transition $\omega_a$ by $\deltaProbe=\omega_c-\omega_a$ (Fig.~\ref{fig:matter-waves}(a)).  After adiabatically eliminating the excited 
state $\ket{e}$ and ignoring mean-field light shifts that will be spin-echoed away, the effective Hamiltonian\cite{VladanOATTheory2010} describing the atom-cavity QND interaction can be expressed in a rotating frame at the atomic transition frequency as 
	\begin{align}
		\hat{\mathcal{H}}_{\text{QND}} = \hbar \left(\delta_c + \chi_{\mathrm{QND}} \hat{\Nu}\right) \hat{c}^{\dagger} \hat{c}
	\end{align}
where the cavity field is described by creation and annihilation operators $\hat{c}^{\dagger}$ and $\hat{c}$. The cavity resonance shifts by an amount $\chi_{\mathrm{QND}}=2 \pi \times \SI{335(4)}{\hertz}$ per atom in $\ket{\uparrow}$ at a detuning $\deltaProbe = 2\pi \times \SI{175}{\mega\hertz}$ (see Methods). The population $\Nu$ of atoms in the momentum state with spin label $\up$ can be estimated by measuring the cavity frequency shift which is estimated by detecting the probe light reflected from the cavity input mirror as the laser frequency is swept across resonance (Figs.~\ref{fig:setup}(a)~and~\ref{fig:squeezing}(a, b)). A typical measurement lasts \SI{150}{\micro\second}. The population $\Nd$ of atoms in the momentum state with spin label $\down$ is measured with the same technique after transferring the atoms to $\up$ using a Raman $\pi$ pulse. The Raman $\pi$ pulse serves the additional functions of re-overlapping the wavepackets and cancelling the average light shift of the probe.

Collective QND measurements are used in creating conditional spin squeezing. 
The spin-momentum projection in the population basis is measured once with the pre-measurement outcome $J_{zp} = \frac{1}{2}\left(\Nu - \Nd\right)|_{\text{pre}}$ , which localizes the state to below the initial coherent spin state level, producing a squeezed state.  The same projection is then measured a second time with the final measurement outcome labeled $J_{zf}=\frac{1}{2}\left(\Nu - \Nd\right)|_{\text{fin}}$. The quantum fluctuation is common to both measurements and can be partially subtracted by considering the difference $\Jzdiff=J_{zf} - J_{zp}$, but any rotation of the state (\textit{i.e.}~ signal) that occurs in the interim appears only in the final measurement outcome. Each final population measurement is made after first optically pumping atoms in $\up$ to $\ket{F=2, m_F=2}$ to achieve lower readout noise (estimated at more than 15~dB below the projection noise level) by using the optical cycling transition to $\ket{F=3, m_F=3}$.

The length of the Bloch vector $\Js$ after the pre-measurement is measured by inserting a $\pi/2$ pulse between the pre- and final measurements (see Methods). Specifically, $\Js$ is estimated from the fringe amplitude of $J_{zf}$ versus the azimuthal phase $\phi$ of the $\pi/2$ pulse as it is varied between 0 to $2\pi$.  The initial length of the Bloch vector $\Jc$ needed for estimating the spectroscopic enhancement is estimated in the same manner, but without the pre-measurement applied. 

Fig.~\ref{fig:squeezing}(c) shows the spectroscopic enhancement $W$ versus the strength of the QND interaction as parameterized by $\Minc$, the average number of incident photons that enter the cavity during each population pre-measurement window. At low $\Minc$, the probe's vacuum noise limits the spectroscopic enhancement, while at high $\Minc$, the spectroscopic enhancement is limited by free space scattering of the probe light that leads to a reduction in $J_{s}$ and transitions to other ground states that decorrelate the pre- and final measurements. Near $\Minc = 600$, $\Nprime = 1170(30)$ atoms, and $\deltaProbe = 2 \pi \times \SI{175}{\mega\hertz}$, we achieve $W=0.46(11)$ or $3.4^{+1.1}_{-0.9}$~dB of directly observed squeezing in the momentum-spin basis.

We also realize entanglement via cavity-mediated interactions \cite{VladanOATExp2010, VladanOATTheory2010, SorensenOATTheory2017}.  The   one-axis twisting (OAT) Hamiltonian \cite{KitagawaUedaOAT}  
	\begin{align}
		\hat{\mathcal{H}}_{\text{OAT}} =\hbar \chi_{\mathrm{OAT}} \hat{J}^2_z\label{eq:oat_hamiltonian}
	\end{align}
is generated by applying a fixed frequency drive tone offset from the average dressed cavity resonance by $\deltaOAT \gtrsim \kappa/2$ \cite{GuoDetuningEnhancedOAT}. Briefly, the populations in each momentum-spin state tune the cavity closer to or further from resonance with the fixed frequency drive tone, allowing more or less light into the cavity such that $\hat{c}^\dagger\hat{c} \propto \hat{N}_\uparrow$.  To first approximation, the spin-light QND Hamiltonian is thus transformed into a spin-only Hamiltonian with a relevant term $\propto \hat{\Nu}^2$. A repeated application of the dynamics after a $\pi$ pulse realizes the Hamiltonian dynamics of Eq.~\ref{eq:oat_hamiltonian}. 

\begin{figure*}[hbt!]
	\centering
	\includegraphics[width=\textwidth]{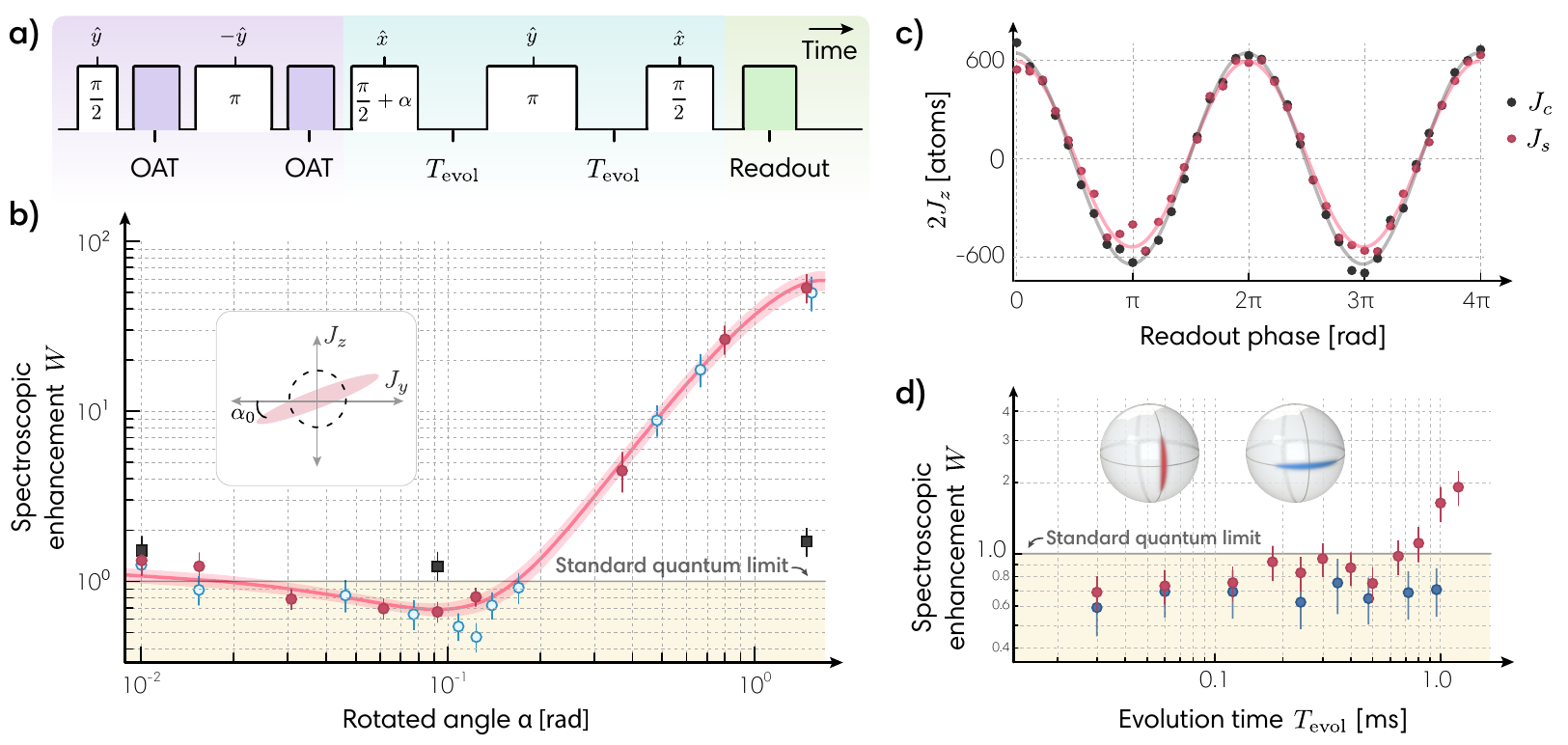}
	\caption{\textbf{Demonstrating sensitivity beyond the standard quantum limit.}
		\textbf{(a)} The squeezed interferometer sequence, including entanglement generation (purple), the interferometer (blue), and state readout (green). Each Raman transition (white) is labeled with magnitude (within) and axis of rotation (above).
		\textbf{(b)} The spectroscopic enhancement $W$ is compared for three configurations: a Mach-Zehnder interferometer with OAT (red circles, sequence above), an unentangled interferometer without OAT (black squares), and OAT-squeezed states without the interferometer (blue circles). The duration of a $\pi/2 + \alpha$ rotation is scanned to minimize the projected spin noise at $\alphanaught$. To model the Gaussian noise distribution, an ellipse is fit with 68\% confidence bands to the OAT-squeezed interferometer data, giving a minimum variance of $W = 0.68(8)$ or $1.7^{+0.5}_{-0.5}$~dB. The interferometer here had $\Tevol = \SI{0.112}{\milli\second}$ and $N = 660(15)$ atoms.
		\textbf{(c)} Interferometer contrast fringes with $\Tevol = \SI{0.112}{\milli\second}$ shown for no squeezing $\Jc$ (black) and with squeezing $\Js$ (red). 
		\textbf{(d)} Phase sensitivity is maintained below the SQL for the squeezed interferometer (red circles, left Bloch sphere) up to $\Tevol = \SI{0.7}{\milli\second}$. By contrast, if the squeezed spin projection is oriented along the population basis (blue circles, right Bloch sphere), spectroscopic enhancement was seen to persist beyond $\Tevol = \SI{1}{\milli\second}$ because this orientation is insensitive to phases accrued during the evolution time. The bias field was $\SI{1}{\gauss}$ along $\hatVert$ for this data. All error bars reported are 1$\sigma$ uncertainties.}
	\label{fig:fringes}
\end{figure*}

The unitary OAT interactions drive shearing of the atomic quantum noise distribution with a resulting squeezed state minimum noise projection oriented at a small angle $\alphanaught$ from $\hat{z}$ ( Fig.~\ref{fig:squeezing}(d) and Fig.~\ref{fig:fringes}(b, inset)).   The state is rotated so that the minimum noise projection is along $\hat{z}$.  The momentum-spin populations are destructively read out as before with measurement outcome labeled $J_{zf}$. The Bloch vector lengths $\Js$ ($\Jc$) with (without) OAT squeezing are also measured just as for the QND squeezing. We directly observe a spectroscopic enhancement from OAT  of $W = 0.56(8)$ or $2.5^{+0.6}_{-0.6}$~dB.  The optimal configuration was realized with $\Minc \approx 700$ photons, $\deltaProbe = 2 \pi \times \SI{350}{\mega\hertz}$, $\deltaOAT = 2.7 \times \kappa/2$, $\chi_{\mathrm{OAT}} \approx 2 \pi \times  10$~Hz and $\Nprime=730(10)$ atoms.

\vspace{2mm}
\noindent\textbf{Entangled matter-wave interferometry.} We now turn to injecting the prepared entangled state into a matter-wave interferometer with the sequence shown in Fig.~\ref{fig:fringes}(a). After preparing a squeezed state with OAT, a Raman beam splitter rotation orients the squeezing along $\hat{y}$. The spin projection $J_y$ will change if a small signal phase $\phi$ is applied.  The orienting of the squeezing is accomplished via a $(\pi/2 + \alphanaught)$ pulse aligned to the atomic Bloch vector along $\hat{x}$. A relative phase accumulates between the wavepackets during a free evolution time $\Tevol$, a Raman $\pi$ ``mirror" pulse is applied, followed by another free evolution time $\Tevol$. Finally, a readout $\pi/2$ pulse transfers the signal $\phi$ and the squeezing into a displacement in the momentum-spin population basis $\hat{z}$ with a measurement outcome $J_{zf}$. The Bloch vector lengths $\Js$ and $\Jc$ are measured in separate experiments with and without OAT applied by scanning the azimuthal phase of the final $\pi/2$ pulse of the interferometer and measuring the fringe amplitude as before (see Fig.~\ref{fig:fringes}(c)).

We achieve a directly observed spectroscopic enhancement  $W= 1.7^{+0.5}_{-0.5}$~dB beyond the standard quantum limit with $\Nprime = 660(15)$ atoms as shown in Fig.~\ref{fig:fringes}(b). Without OAT, the performance of our interferometer is worse than the SQL due to imperfect interferometer contrast $C_i = 2 \Jc / \Ninit \approx 0.9$. We note that the actual phase variance of the squeezed interferometer is improved by $3.4^{+0.9}_{-1.2}$~dB compared to this unsqueezed interferometer (see Methods).

Phase sensitivity beyond the SQL was limited to evolution times $\Tevol < \SI{0.7}{\milli\second}$ (Fig.~\ref{fig:fringes}(d)). A comparable level of decrease in phase sensitivity was observed in an identical sequence in which all optical Raman pulses were replaced by equivalent microwave pulses suggesting that the spin degrees of freedom may be responsible for the observed loss in sensitivity. We also observe that if the squeezed spin projection is left in the population basis $J_{z}$ during the interferometer, then the squeezing persists for several milliseconds.  From this, we conclude that the entangled state persists for longer than we can directly confirm.

In the future, the combination of Raman and Bragg techniques demonstrated here would enable the most delicate portion of the interferometer to be operated fully with the two portions of the superposition possessing the same spin label.  

To further improve interferometer sensitivity, the entanglement can be combined with large momentum transfer sequences or one could inject the squeezed state into a lattice interferometer to hold the atoms for longer \cite{MullerProbingGravity20Sec}.  One could also prepare the entanglement in the cavity and allow the atoms to undergo free fall outside of the cavity with readout via fluorescence measurement  \cite{KasevichSqueezedClock2020}, another promising path for scaling to larger momentum transfers and longer interferometer times. The amount of momentum squeezing could be improved with larger collective cooperativity $N \Coop$.  The need for velocity selection limits our final number of atoms, so higher atom density in momentum space through improved axial cooling or the use of a Bose-Einstein condensate could lead to significant improvements \cite{DanaBECChipInterferometer, KetterleBECINterferometerSqueezing, OberthalerSqueezedBECInt, RaselTwinLatticeAI2021}. As the atom number is increased, it will be necessary to reduce the level of classical rotation-added noise or to make the added noise common mode as is done for gravity gradiometers and for proposed gravity wave and dark matter detectors \cite{RajendranGravityWaveSensor2013,ThompsonGravWaveDetection2017,MIGAInterferometer2018,MAGIS100DarkMatterDescription2021}.

In this work, the OAT-squeezed states were successfully used to realize a squeezed matter-wave interferometer, whereas the QND-squeezed states were not. The OAT produced states were generated at lower atom number and associated smaller momentum spread, leading to less classical added rotation noise relative to the SQL and reduced shortening of the Bloch vector during the rotations. The QND-squeezed states would be enhanced by improving the total effective quantum efficiency from $q\approx0.1$ here, with for instance $q\approx0.4$ in previous work\cite{Thompson17dBSqueezing}.

It may also be possible to generate spin-squeezed states utilizing optical cycling transitions in rubidium, strontium, and ytterbium \cite{ZilongsPRA, Thompson10dBSqueezing, Thompson17dBSqueezing, ThompsonHomogeneousCoupling, VuleticSqueezedOpticalClock2020} and then use Raman transitions to map the entanglement to purely momentum states  \cite{MullerSDK2018,KlemptMomentumEntanglementArxivXXX}.  The fundamental scaling of the achievable Wineland parameter would improve to $W\propto 1/N \Coop$ from the current scaling $W\propto 1 / \sqrt{N \Coop}$ \cite{ZilongsPRA}. Indeed, the combination of larger atom number and probing on a cycling transition are the primary reasons for the larger amounts of squeezing achieved in previous work \cite{Thompson17dBSqueezing,KasevichSqueezing2016} compared to the present results.

This proof-of-principle light-pulse matter-wave interferometer paves the way for utilizing cavity-generated entanglement as a quantum resource, enabling the next generation of interferometers with higher precision, enhanced measurement bandwidth, higher accuracy, and smaller size. Such devices will advance the frontiers of both practical applications and discoveries in fundamental science.

\begin{acknowledgments}
We acknowledge funding support from the National Science Foundation under Grant Numbers 1734006 (Physics Frontier Center) and  OMA-2016244 (QLCI), DOE Quantum Systems Accelerator, NIST, and DARPA.  We acknowledge helpful feedback on the manuscript from Dana Z. Anderson and Ana Maria Rey, helpful discussions with Matthew Jaffe and Nicola Poli, and laser locking development by Denton Wu.
\end{acknowledgments}

\bibliography{entanglement_enhanced_interferometer}

\clearpage
{\noindent\bfseries\Large{Methods}}

\section{Blue-detuned donut dipole guide}
The blue dipole guide laser is a $760$~nm interference filter ECDL locked to a reference cavity for improved long-term stability. The laser is modulated by a fiber EOM with modulation index $\beta \approx 1.3$ at the cavity free spectral range $\omegaFSR$.  By exciting adjacent longitudinal modes of the cavity with opposite spatial parity with respect to the center of the cavity, one creates an axially-uniform blue dipole guide near the center of the cavity \cite{ThompsonHomogeneousCoupling}. The donut-mode LG$_{01}$ profile is constructed from the $\pm 1$st diffraction orders of a fork-pattern phase plate. Stress-induced birefringence of the cavity mirrors breaks cylindrical symmetry and splits the Hermite-Gaussian HG$_{10}$ and HG$_{01}$ modes up to $\delta_{\text{HG}} = 2 \pi \times \SI{100}{}-\SI{500}{\kilo\hertz}$, depending on cavity piezo voltage, to be compared to the 157(5)~kHz FWHM cavity linewidth for these modes. For the data presented here, $\delta_{\text{HG}} = 2 \pi \times 350 $~kHz. Prior to entering the cavity, the two LG modes are sent along separate paths. One path enters a free-space EOM to generate sidebands for locking the cavity to the blue dipole guide laser. The other path passes through two AOMs with a $\delta_{\text{HG}}$ frequency difference such that the projected HG modes combine within the cavity to approximate an LG$_{01}$ mode's radial intensity distribution via LG$_{01} = \textrm{HG}_{01} + i \textrm{HG}_{10}$. Because the frequency splitting $\delta_{\text{HG}}$ is much greater than the radial trap frequency, the atoms effectively experience the time-averaged radial trapping potential of an LG$_{01}$ mode.

\section{Laser cooling}
The experimental sequence is repeated every 750~ms.  Each trial begins with a 2D MOT loading a 3D MOT with $10^{8}$ atoms near the cavity center for approximately \SI{0.5}{\second}. The MOT coils are turned off, and around $2\times 10^{5}$ atoms are cooled via polarization gradient cooling to \SI{15}{\micro\kelvin} and loaded into an 813.5~nm red-detuned intracavity lattice with FWHM cavity linewidth 166(5)~kHz. Additional radial confinement is provided by the blue dipole guide. The red lattice depth is ramped down to a depth of \SI{80}{\micro\kelvin} or $ 250 E_l$ where $E_l$ is the recoil energy of the lattice. We then apply $\Lambda$-enhanced grey molasses cooling. Each of the six molasses beams has \SI{2.5}{\milli\watt} and \SI{1}{\centi\meter} beam waist. The light is detuned $2 \pi \times \SI{42}{\mega\hertz}$ blue of $\ket{F=2} \rightarrow \ket{F'=2}$. A fiber EOM generates a \SI{100}{\micro\watt} sideband for coherently forming the $\Lambda$ system as $\ket{F=1} \leftrightarrow \ket{F'=2} \leftrightarrow \ket{F=2}$. After \SI{5}{\milli\second}, the temperature of the ensemble is reduced to \SI{6}{\micro\kelvin}.

We then perform two-dimensional degenerate Raman sideband cooling (RSBC) to further cool the radial temperature \cite{ChuRamanSidebandCoolingCs1998}. Three RSBC beams form a triangular lattice in a plane perpendicular to the cavity axis $\hatVert$, with trapping frequency $\omegatri = 2 \pi \times \SI{75}{\kilo\hertz}$. The blue dipole guide and red lattice continue to provide a background radial trap. The RSBC laser is blue-detuned \SI{50}{\giga\hertz} from the $\ket{F=1} \leftrightarrow \ket{F'=2}$ transition so that atoms are trapped at the nodes of the triangular lattice, suppressing scattering off the cooling beams. A bias magnetic field of \SI{0.11}{\gauss} along $\hatVert$ is applied to match the first-order Zeeman splitting to the trap frequency $\omegatri$. The polarizations of the three beams are twisted by $10^{\circ}$ from vertical to create the Raman coupling for driving the vibrational mode transition $\ket{F=1, m_F, \ntri}\rightarrow\ket{F=1, m_F-1, \ntri-1}$ that reduce the vibrational quantum number $\ntri$ in the local traps. During RSBC, atoms are continuously repumped back to $\ket{F=1, m_F=1}$ by a separate laser.

To improve the coupling of the atoms to the cavity we apply multiple cooling cycles each lasting 2~ms. The RSBC light is ramped on over 0.3~ms, cooling occurs for 1.2~ms, and then RSBC light is ramped off over 0.3~ms. After \SI{225}{\micro\second}, the atoms have oscillated back to the center of the cavity, at which point we repeat the cooling cycle. After three cooling cycles, we slowly turn off the remaining red lattice and the RSBC lattice in 3~ms so that the atoms start to free fall. Atoms are then optically-pumped to $\up$ using a pair of $\pi$-polarized laser beams on resonance with the $\ket{F=1} \rightarrow \ket{F'=2}$ and $\ket{F=2} \rightarrow \ket{F'=2}$ transitions applied transverse to the cavity axis.  Just before the interferometer sequence, the radial temperature is \SI{1.4(5)}{\micro\kelvin}. State transfer with microwave pulse may be employed for future improvement to reduce heating associated with optical pumping.

\section{Atomic and cavity probe lasers}
To stabilize the frequencies of the Raman lasers and the atomic probe relative to the cavity, we frequency lock a separate cavity probe laser to the cavity and then perform offset frequency phase locks to this laser.  The cavity probe is locked to a cavity TEM$_{00}$ mode approximately \SI{160}{\giga\hertz} to the blue of the atomic transition frequency $\omega_{a}$ such that this mode is essentially unperturbed by the presence of atoms.  The locking of the cavity probe to the cavity is done via a Pound-Drever-Hall lock at very low phase modulation index for a single sideband to carrier power ratio of $10^{-4}$.  Rather than locking to the carrier, we lock to the weak sideband.  This allows us to reduce the amount of power entering the cavity to only  \SI{400}{\pico\watt} (half from the sideband and half nonresonantly from the carrier) while still operating above the technical noise floor of the photodiode. This lock is always engaged.  To allow phase locking of other lasers to the cavity probe with relative beat notes less than $\SI{2}{\giga\hertz}$, some of the laser light is passed through a fiber EOM driven strongly at \SI{13.6}{GHz} to generate very high order sidebands.

The atomic probe laser is phase-locked with an offset frequency of approximately $13.6\times12=\SI{163.2}{\giga\hertz}$ to the red of the cavity probe, placing it close to $\omega_a$. The offset phase-lock frequency is adjusted to maintain the atomic probe laser approximately $\delta_c/2\pi+80$~MHz blue of $\omega_a$. We derive three important tones from this laser:  a homodyne reference beam, a path length stabilization beam used for removing path length noise and drift, and the actual atomic probe tone used for one axis twisting and QND measurements.  The path length stabilization beam is passed through an EOM that is modulated at 80~MHz to create a weak sideband that will serve as the atomic probe tone.  The combined path length stabilization and atomic probe tones are reflected from the cavity and detected on a single homodyne detector.  The homodyne reference beam is shifted by an $80$~MHz AOM to have the same frequency as the atomic probe tone.  The quadrature of the atomic probe tone that we detect in homodyne is actively stabilized by adjusting the phase of the homodyne reference tone.  This is achieved by detecting the phase of the path length stabilization tone appearing in the homodyne detector at 80~MHz  and then holding this phase constant by feedback on the frequency of the 80~MHz AOM used to shift the homodyne reference beam.

The laser could be actively locked to the dressed resonance as in \cite{Thompson17dBSqueezing}, or the linear part of the dispersive could be used to estimate small frequency shifts, but for this work, we sweep the atomic probe laser,  and all derived beams, so that the atomic probe tone sweeps through cavity resonance resonance at \SI{1.5}{\mega\hertz/\milli\second}. Although this simplifies the experiment, it results in a \SI{6}{\decibel} loss of quantum efficiency for a fixed amount of free space scattering when compared to performing homodyne detection on line center.  Including this loss of efficiency, the net effective quantum efficiency is approximately 10\%.

When using the atomic probe to drive one-axis twisting, it is ideal to operate with the driving laser detuned from cavity resonance by $\deltaOAT = \kappa/2$ to suppress free-space scattering. However, we work at larger detunings for two reasons.  First, an increased detuning reduces deleterious QND interactions (or equivalently, photon shot noise from the applied drive tone) that were neglected in our description of the emergence of the unitary dynamics \cite{GuoDetuningEnhancedOAT}. Secondly, this allows operation in a linearized regime even in the presence of shot-to-shot total atom number fluctuations.  We empirically find an optimum detuning of  $\deltaOAT = 2.7 \times \kappa/2$ with $\chi_{\mathrm{OAT}} \approx 2 \pi \times  10$~Hz.

The cavity probe, atomic probe and Raman lasers are DBRs with free-running linewidths of approximately 500~kHz.  We use external optical feedback to narrow their linewidths  \cite{ VladanNarrowDBR2}. A small fraction of the power from each laser is picked off and then retro-reflected back into the laser with a round trip length of 3~m in free space.  The frequency of each laser is primarily determined by the length of the optical feedback path length which is stabilized using a piezo to move the retro-reflection mirror and a free-space phase modulator EOM for fast actuation with unity gain frequency of 500~kHz. By optimizing the optical feedback fraction typically between $10^{-4}$ to $10^{-3}$, we achieve Lorentzian linewidths of less than \SI{1}{\kilo\hertz}.

\section{Microwave source}
High fidelity Raman pulse sequences require agile control of low-phase noise microwaves. Our microwave source is based on Ref.~ \cite{ThompsonMicrowaveSource}. A low phase noise \SI{100}{\mega\hertz} crystal oscillator (Wenzel ULN 501-16843) is multiplied to \SI{6.800}{\giga\hertz} using a nonlinear transmission line frequency comb generator (Picosecond Pulse
Labs LPN7110-SMT). The stable 6.800~GHz is provided as the local oscillator for a single sideband modulator (Analog Devices HMC496).

The required $I$ and $Q$ modulation inputs to the single sideband modulator are created using three RF tones from an Analog Devices AD9959 DDS. Two RF tones are at the same frequency near  135~MHz and are 90 degrees out of phase.  The phase, frequency, and amplitude of these two tones can be jumped for arbitrary rotations on the Bloch sphere, for selecting different momentum-changing transitions, velocimetry, etc.  The third RF tone starts near 100~MHz but is continuously ramped in frequency at a rate $2 k g \approx 2\pi \times \SI{25.1}{\kilo\Hz/\milli\s}$ to match the time variation of the two-photon Doppler shift as the atoms fall under gravity.  Each of the two initial RF tones are mixed with this third signal to generate tones near 35~MHz for the $I$ and $Q$ inputs to the single sideband modulator.

Finally, the modulator output near 6.835~GHz is divided by two using a low-noise divider (Analog Devices HMC862A) and applied to a fiber-coupled EOM to generate the desired Raman tones as the $\pm 1$st order sidebands. We estimate that the noise contributed by this frequency source is at least \SI{30}{\decibel} below the SQL for 1000 atoms.


\section{Raman transitions and velocity selection}
The laser that drives the Raman transitions is detuned $\Delta = 2 \pi \times \SI{85}{\giga\Hz}$ blue of $\omega_a$. As is done for the atomic probe, the Raman laser is stabilized with respect to the cavity by an offset frequency phase lock to the cavity probe.  The offset frequency is set to center the Raman laser between two adjacent longitudinal TEM$_{00}$ cavity modes. The two Raman tones, whose generation is described above, are symmetrically detuned from the cavity resonances by approximately $(\omegaHF - \omegaFSR)/2 = 2\pi \times \SI{23}{\mega\Hz}$.   With \SI{2.5}{\milli\watt} of total $\sigma^{+}$-polarized light incident on the cavity, the EOM modulation index allows a maximum observed two-photon Rabi frequency of $\OmegaTwoPh = 2 \pi \times \SI{15}{\kilo\hertz}$, with the Rabi frequency tuned to smaller values by adjusting the total incident power using an AOM.  For the large momentum transfers shown in Fig.~\ref{fig:matter-waves}(e), Bragg transitions are driven by two laser tones derived from the same laser with difference frequency $\omega_B = \deltaVS - b \left(t-\TVS\right)$, where $b$ is the chirp rate defined below.

As atoms fall under gravity, the relative Doppler shift for light propagating upwards versus downwards chirps linearly in time.  We compensate this effect by linearly ramping the instantaneous frequency of the sidebands as $2 \omega_R = \omegaHF + \delta - b (t-\TVS)$ with chirp rate  $b=2 \pi \times \SI{25.11}{\kilo\hertz / \milli\second} \approx 2kg_\parallel$.  Here $g_\parallel=9.8 \mathrm{m/s^2}$ is the projection onto the cavity axis of the local acceleration due to gravity, $\delta$ is the two-photon detuning in the falling frame of reference, and $\TVS$ is the time at which we will apply the first $\pi$ pulse for velocity selection described below.  We also note that during rotation pulses, we adjust the two-photon detuning $\delta$ by approximately 4~kHz (in a phase coherent manner) to compensate for differential AC Stark shifts of the two pseudo-spin states induced by the Raman beams.  In the accelerating reference frame, the phase of the interferometer fringe evolves quadratically with $T$ nominally as $\phi = (2 k g_\parallel - b) T^2$ from which we extract a value of $g_\parallel$ consistent with the known local value of gravitational acceleration to within the uncertainty of the angular orientation of the cavity axis with respect to local gravity.  The chirp rate $b$ is nominally tuned so that $\ll 1$~rad of phase evolves in the accelerating frame, but in the lab frame, the accumulated phase evolves as $\phi = 2 k g_\parallel T^2$, approximately $2500$~rad for the largest $T=4$~ms explored in this work.

After being released from the 813~nm  lattice and falling for $\Tfall=15$~ms, the atoms are optically-pumped to $\up$, and the two-photon Raman detuning is set to  $\deltaVS= - 2\pi \times \SI{400}{\kilo\hertz} \approx b \Tfall$  to transfer a group of atoms to $\down$ from the center of the axial velocity distribution \cite{ChuVelocitySelection}. Atoms in $\up$ are removed by a transverse radiation pressure force. The velocity selection is then repeated to further narrow the momentum width of the selected atoms down to $\Delta p < 0.1 \hbar k$ set by the two-photon Rabi frequency $\OmegaTwoPh = 2\pi \times \SI{1.4}{\kilo\hertz}$.

The Raman laser is a DBR laser with a free-running linewidth of approximately 500~kHz. We observed that the cavity converted laser frequency noise to intracavity amplitude noise near $\deltaVS$ that can resonantly drive undesired Bragg transitions, leading to a loss of nearly 50\% of population to other momentum states outside of the desired two-level basis for all the Raman pulses involved in the interferometer sequence combined. We note that in the symmetric detuning configuration here, the Raman transitions are first order insensitive to conversion of laser frequency noise to both AM and PM noise on the intracavity Raman tones.  However, the Bragg transitions are first order sensitive because of the opposite parity of the standing wave modes being driven.

After narrowing the laser to a Lorentzian linewidth of less than \SI{1}{\kilo\hertz}, we found the fraction of total atoms lost out of the desired two-level manifold is less than 3(3)\% for all the Raman pulses involved in the interferometer sequence combined. We also observed residual off-resonance transitions to other momentum states if the turn on and off of the Raman beams was too rapid.  The fraction of atoms lost per pulse was reduced to 0.2(1.0)\% per pulse by using an rf switch with \SI{3}{\micro\second} rise-time to gate the Raman tones. Without the shortening of the Bloch vector $\Jc$ from the two effects, we estimate that the observed spectroscopic enhancement could be improved by 0.2(2)~dB for the full interferometer sequence.

\section{Wineland criterion}
The Wineland criterion is often presented in the following form \cite{Thompson17dBSqueezing}
	\begin{align}
		W =\frac{\left(\Delta J_{z} \right)^2 C_i}{\Delta J_{z, SQL}^2 C_f^2},
	\end{align}
where the contrasts  are related to Bloch vector lengths here by $C_i \equiv 2 \Jc / \Ninit$ and $C_f \equiv 2 \Js / \Ninit$ for  total atom number $\Ninit$. By rearranging terms, it can also be expressed in a more physically meaningful form as the ratio $W=\left( \Delta \theta  / \Delta \thetaSQLi \right)^2$ between the observed angular resolution $\Delta \theta= \frac{ \Delta J_z  }{ \Js } $ with entanglement and the standard quantum limit $ \Delta \thetaSQLi = 1 / \sqrt{N} \equiv 1 / \sqrt{2\Jc}$ for a pure state with the same Bloch vector length $\Jc$ as that of the actual mixed state when entanglement is not created.  

We now establish the connection between the spin operators and actual experimental measurements. We define the cavity frequency shifts induced by a single atom in $\ket{F=2,m_F=2}, \ket{F=2,m_F=0}$ and $\ket{F=1,m_F=0}$ as $\chi_2,\chi_0\equiv\chi_{\mathrm{QND}}$ and $\chi_\downarrow$ respectively. 

For OAT squeezing, we estimate the angular resolution $\Delta\theta$ after the squeezing generation or the full squeezed interferometer sequence as follows. To measure the final spin projection $J_{zf}$, we optically pump the atoms in $\up$ to $\ket{F=2,m_F=2}$, measure the cavity frequency shift with outcome labeled $\omega_{1f}$, blow away atoms in $\ket{F=2}$, apply a Raman $\pi$ pulse, optically pump the atoms in $\up$ to $\ket{F=2,m_F=2}$, and measure a second cavity frequency shift with outcome labeled $\omega_{2f}$. We estimate the final spin projection $J_{zf}$ from the difference between the two cavity frequency shifts  $J_{zf}=\frac{\omega_{1f}-\omega_{2f}}{2\chi_2}-\frac{\epsilon}{\chi_2} \omega_{2f}$, where $\epsilon=\frac{\chi_\downarrow/2}{\chi_2}$. To convert the spin projection $J_{zf}$ into an estimate of the Bloch vector polar angle $\theta_f$, we measure the length of the Bloch vector $J_s$ by scanning the azimuthal phase $\phi$ of the readout $\pi/2$ pulse. In the case of the squeezed interferometer, this is the final $\pi/2$ pulse of the interferometer and just prior to the measurement $J_{zf}$. In the case of OAT squeezed state generation, this is an added $\pi/2$ pulse after the squeezing and just prior to the measurement $J_{zf}$. We fit the resulting differential cavity frequency shifts $\left(\omega_{1f}-\omega_{2f} \right)|_\phi$ to the function $y_0+A_f \sin \left(\phi-\phi_0\right)$. The Bloch vector length is then estimated by $J_s=\frac{A_f}{2\chi_2-\chi_\downarrow}$. The Bloch vector polar angle $\theta_f$ from the final measurement is thus estimated by $\theta_f = \frac{J_{zf}}{J_s} = \frac{ \omega_{1f} - \omega_{2f}}{A_f} - \epsilon \frac{\omega_{1f}+\omega_{2f}}{A_f} + 2 \epsilon^2 \frac{\omega_{2f}}{A_f}$. The angular resolution  $\Delta\theta$ is approximated as $\Delta\theta=\Delta\theta_f \approx \frac{\Delta \left( \omega_{1f}-\omega_{2f} \right)}{A_f}$, where we note the scale factors $\chi_2$ etc. are canceled at the order of $\epsilon^0$. With a typical $\lvert \epsilon \lvert<1/50$ and the fractional total number fluctuation $\Delta \left(\frac{\omega_{1f}+\omega_{2f}}{A_f}\right)$ being less than $0.03$, the corrections of order $\epsilon^1$ would need to be included for squeezing 30~dB below the SQL.

For the QND measurements, we perform pre-measurements to localize the quantum state and use the final measurements to verify the squeezing generated by the pre-measurements as described before. The phase resolution is defined as the phase fluctuation between the pre- and final measurements $\Delta\theta=\Delta\left(\theta_p-\theta_f\right)$. The Bloch vector polar angle of the final measurements $\theta_f$ is estimated as in the OAT measurement with the atomic population optically pumped to $\ket{F=2,m_F=2}$. For the pre-measurements, we measure pairs of cavity frequency shifts $\omega_{1p}$ and $\omega_{2p}$ separated by $\pi$ pulses but without the optical pumping so the atomic population is in $\up$ during the cavity frequency shift measurements. The spin projection $J_{zp}$ in the pre-measurements is estimated from the differential frequency shift $J_{zp}=\frac{\omega_{1p}-\omega_{2p}}{2\left(\chi_0-\chi_\downarrow\right)}$. The length of the Bloch vector $J_s$ just after the pre-measurement is measured by adding a $\pi/2$ pulse just after the pre-measurement and scanning its azimuthal phase $\phi$, after which we perform a single cavity frequency shift measurement with outcome labeled $\omega_{1f}|_\phi$. We then fit the resulting fringe to the function $y_0+A_p \sin \left(\phi-\phi_0\right)$ and estimate the Bloch vector length $J_s=\frac{A_p}{\chi_0-\chi_\downarrow}$. The Bloch vector polar angle $\theta_p$ is evaluated $\theta_p = \frac{J_{zp}}{J_p} = \frac{\omega_{1p}-\omega_{2p}}{2 A_p}$. As before, the angular phase resolution is sufficiently approximated by keeping only to the order of $\epsilon^0$ as $\Delta\theta=\Delta\left(\theta_p-\theta_f\right)\approx \Delta \left( \frac{\omega_{1p}-\omega_{2p}}{2A_p} - \frac{\omega_{1f}-\omega_{2f}}{A_f} \right)$ with no dependence on scale factors $\chi_2$, $\chi_0$ or $\chi_\downarrow$.

For estimating the standard quantum limit $\Delta\theta_{SQL}$, we measure the length of the Bloch vector $J_c=J_s|_{M_i=0}=\frac{{A_p}|_{M_i=0}}{ \chi_{0}-\chi_{\downarrow}} $ using the same sequence for measuring $J_s$ in the QND pre-measurements described just above but setting the photon number $M_i$ to zero during the pre-measurements or squeezing for QND measurement or OAT respectively. To estimate the standard quantum limit $\Delta \theta_{SQL} = 1/ \sqrt{N}=1 / \sqrt{2 J_c}$ we therefore need to know accurate values of $\chi_0$ and $\chi_\downarrow$. To sufficient approximation $\chi_{0}  = g^2 \left( \frac{B_3 }{\delta_c} + \frac{B_2 }{\delta_c + \delta_2} + \frac{B_1}{\delta_c + \delta_1} \right)$ with atom-cavity coupling $g$ discussed below, hyperfine splittings $\delta_2= 2 \pi \times 266.7~\mathrm{MHz}, \delta_1= 2 \pi \times 423.6~\mathrm{MHz}$ and branching ratios $B_3= \frac{6}{15}, B_2=\frac{3}{12}, B_1=\frac{1}{60}$ of the excited states $\ket{F'=3,2,1, m_F=1}$ to the ground state $\up$ transition that interact with the probe light. To sufficient approximation $\chi_{\downarrow}=g^2 \left( \frac{B_{2,\downarrow} }{\delta_c + \delta_2 - \omegaHF} + \frac{B_{1,\downarrow}}{\delta_c + \delta_1 - \omegaHF} \right)$ with branching ratios $B_{2,\downarrow}=\frac{3}{12}, B_{1,\downarrow}=\frac{5}{12}$ of the $\ket{F'=2,1, m_F=1}$ to the ground state $\down$ transition. Though not used in the calculation, the cavity frequency shift from a single atom in $\ket{F=2,m_F=2}$ is approximated by $ \chi_2 = \frac{g^2} {\delta_c} $ for the cycling transition between the excited state $\ket{F=3,m_F=3}$ and ground state $\ket{F=2,m_F=2}$. 

The maximum single-atom vacuum Rabi splitting $2 g_0=2 \sqrt{\frac{2  D^2 \omega_c} {\pi L w_0^2 \epsilon_0 \hbar}} =2\times 2\pi \times  0.4853(5)~\mathrm{MHz}$ \cite{VladanQNDClock2010} with fractional uncertainty dominated by the fractional uncertainty ($1.1\times 10^{-3}$) on the dipole matrix element $D$ for the $\ket{F=2,m_F=2}$ to $\ket{F=3,m_F=3}$ transition, and $\epsilon_0$ the vacuum permeability. The cavity length $L$ and mode waist $w_0$ are determined very precisely by measuring the free spectral range and transverse mode frequency splitting. Since the atoms traverse many standing waves of the cavity during the measurement windows, we can coarse grain over the standing waves to arrive at a time-averaged spatially dependent coupling $g_t\left(r,z\right) = \frac{g_0}{\sqrt{2}} \frac{e^{-r^2/w_0^2}}{ \sqrt{ 1+\left( \frac{z}{Z_r} \right)^2 } }$, where $Z_r=2.1$~cm is the Rayleigh range of the cavity  \cite{ThompsonHomogeneousCoupling}. The effective single atom-cavity coupling frequency is given by the ensemble averaged moments of the spatially dependent $g_t\left(r,z\right)$ as $g=\sqrt{ \frac{\braket{g_t\left(r,z\right)^4}}{\braket{g_t\left(r,z\right)^2}} } = \frac{g_0}{\sqrt{2}} \left( 1- f_{cor} \right)= 2\pi \times0.341(2)~\mathrm{MHz}$ \cite{VladanQNDClock2010}. The final fractional uncertainty ($6\times10^{-3}$) on $g$ is dominated by the uncertainty on the correction factor $f_{cor}\approx \frac{z_0^2+\sigma_z^2}{2 Z_r^2} + \frac{r_{rms}^2}{w_0^2}$, where $z_0=1\left(2\right)$~mm is the axial position of the cloud relative to the cavity center, $\sigma_z=0.5 (3)$~mm is the RMS axial spread of the cloud, and $r_{rms}$ is the RMS cloud radius of the atoms. The fractional uncertainty on $g$ contributed from $z_0$, $\sigma_z$ and  $r_{rms}$ are $5\times 10^{-3}$, $4\times 10^{-3}$ and $2\times 10^{-3}$ respectively.

The uncertainties on the cavity detuning $\delta_c = 175(2)$~MHz or $350(2)$~MHz lead to fractional uncertainties $\le 0.01$ on $\left( \chi_{\mathrm{QND}} - \chi_{\downarrow} \right)$. Because the atoms move along the cavity axis, the probe light is Doppler shifted by of order $\deltaVS/2$; however, here $\delta_c \gg \deltaVS$ so that there is only a negligible fractional correction to $\chi_{\mathrm{QND}}$ of order $(\deltaVS/2 \delta_c)^2\lesssim 10^{-6}$. The effect of spread in momentum states is even more negligible.

Combining uncertainties from $g$ and $\delta_c$, the fractional uncertainty on $\left( \chi_{\mathrm{QND}} - \chi_{\downarrow} \right)$ is $\le 1.4\times10^{-2}$. This uncertainty combined with the fractional uncertainty on the fitted fringe amplitude $A_p$ of $9\times10^{-3}$ yields a total fractional uncertainty on the standard quantum limit variance $\left(\Delta \thetaSQL\right)^2 $ of $1.7\times10^{-2}$. To estimate the angular resolutions $\left( \Delta\theta \right)^2$, we typically use 100 to 200 experimental trials, which leads to a typical statistical fractional uncertainty on $\left( \Delta\theta \right)^2$ of 0.1 to 0.2. The final reported uncertainties on the Wineland parameters are thus dominated by the statistical uncertainties on the phase resolution $\left( \Delta\theta \right)^2$.

Without the QND pre-measurements or one-axis twisting, the mixed state actually performs worse than the standard quantum limit, conceptually due to the spin noise from the dephased or decohered fraction of the atoms that contribute noise but no signal.  This is why the observed improvement in the interferometer sensitivity is larger than the Wineland parameter; however, the Wineland parameter captures what fraction of the improvement can be certified to arise due to entanglement between the atoms and not due to just cancellation of spin noise alone.

\section{Vibration noise}

Mechanical vibrations of the cavity mirrors are equivalent to a fluctuating phase reference for the atoms. A commercial vibrometer was used to measure the spectral density $S_a(\omega)$ of acceleration noise at a location on the optical table close to the portion that supports the vacuum chamber. In the limit of zero-duration pulses, the transfer function for a Mach-Zehnder interferometer $\lvert T(\omega)\rvert^2=\frac{64 k^2}{\omega^4} \sin\left({\frac{\omega \Tevol}{2}}\right)^4$ converts accelerations to an integrated phase noise $\phi^2 = \int_{0}^{\infty} |T(\omega)|^2 S_a(\omega) \, d\omega$. For a sequence with $\Tevol = \SI{0.3}{\milli\second}$, we estimate the phase noise caused by vibrations is 20~dB lower than the phase resolution set by the SQL of 1000 atoms.

\section{Author contributions}

G.P.G., C.L., and B.W. contributed to the building of the experiment, G.P.G. and C.L. conducted the experiments and data analysis. J.K.T. conceived and supervised the experiments. G.P.G., C.L., and J.K.T. wrote the manuscript. All authors discussed the experiment implementation and results and contributed to the manuscript.

\section{Data availability}

All data obtained in the study are available from the corresponding author upon reasonable request.

\section{Competing interests}

The authors declare no competing interests.

\clearpage

\end{document}